\documentclass[preprint,showkeys,superscriptaddress,amsmath,amssymb]{revtex4-1}

\usepackage{times}
\usepackage{graphicx}
\usepackage{dcolumn}
\usepackage{amsmath}
\usepackage[dvipsnames]{xcolor}
\usepackage{setspace}

\usepackage{amssymb}
\usepackage[version=3]{mhchem}

\usepackage{upgreek}
\usepackage{textcomp}

\usepackage{multibib}

\begin{document}


\title{{Modeling metasurfaces using discrete-space impulse response technique}}

\author{Mahsa Torfeh}
\affiliation{Department of Electrical and Computer Engineering, University of Massachusetts Amherst, 151 Holdsworth Way, Amherst, MA 01003, USA}
\author{Amir Arbabi}
\email{arbabi@umass.edu}
\affiliation{Department of Electrical and Computer Engineering, University of Massachusetts Amherst, 151 Holdsworth Way, Amherst, MA 01003, USA}

\maketitle
{\section*{Abstract}}
Metasurfaces are arrays of subwavelength meta-atoms that shape waves in a compact and planar form factor. Analysis and design of metasurfaces require methods for modeling their interactions with waves. Conventional modeling techniques assume that metasurfaces are locally periodic structures excited by plane waves, restricting their applicability to gradually varying metasurfaces that are illuminated with plane waves. Here we introduce the discrete-space impulse response concept that enables the development of accurate and general models for metasurfaces. According to the proposed model, discrete impulse responses are assigned to metasurface unit cells and are used to determine the metasurface response to any arbitrary incident waves. We verify the accuracy of the model by comparing its results with full-wave simulations. The proposed concept and modeling technique are applicable to linear metasurfaces with arbitrary meta-atoms, and the resulting system-level models can be used to accurately incorporate metasurfaces into simulation and design tools that use wave or ray optics.\\

{\section*{Introduction}}

Metasurfaces are rationally designed subwavelength arrays of meta-atoms that perform various transformations on incident waves.  In contrast to conventional optical components that operate based on reflection and refraction, metasurface components utilize scattering from subwavelength meta-atoms and achieve unprecedented control over the phase, amplitude, and polarization of transmitted or reflected waves \cite{1-Neshev2018,2-Kamali2018,3-Chen2016,4-Chang2018,5-Scheuer2017,6-Jahani2016,7-Kildishev2013}. Accurate full-wave modeling of metasurfaces is challenging and, in most cases, unfeasible because typical metasurfaces are significantly larger than their operation wavelength and have subwavelength features. Therefore, approximate methods are commonly used for their design and analysis. Aperiodic metasurfaces with periodic lattices are a significant category of metasurfaces because of the ease of their designs, and a large number of components based on such metasurfaces have been demonstrated at microwave and optical frequencies \cite{1-Neshev2018,2-Kamali2018,8-Zhan2017,9-Liu2018,10-Horie2018,11-Khorasaninejad2017,12-Ang2017,13-Aieta2012}. 

Figure~\ref{fig:FigOneLabel}a shows a schematic illustration of an aperiodic metasurface with a periodic lattice. The conventional approach for modeling such metasurfaces relies on the local periodicity assumption. If the meta-atoms’ geometries vary slowly enough from one unit cell of a metasurface to the next, then the metasurface may be considered approximately periodic in the neighborhood of each meta-atom. Thus, the interaction of an incident plane wave with the metasurface may be modeled by gradually varying transmission or reflection coefficients whose values at each unit cell depend only on the meta-atom inside that unit cell.  The local complex-valued transmission coefficient corresponding to each meta-atom is then approximated by the transmission coefficient of a periodic array composed of the same meta-atom at normal incidence (Fig.~\ref{fig:FigOneLabel}b). Because the transmission coefficients of periodic arrays of meta-atoms can be computed quickly using Fourier modal techniques or by full-wave simulations of one of their unit cells, the computational cost of the conventional method is low. Thus the approach is suitable for analyzing large structures and has been extensively used in the design and modeling of metasurfaces \cite{14-Arbabi2015,15-Arbabi2015b,16-Vo2014,17-Devlin2016}. 

\setcounter{figure}{0}
\begin{figure}[t]
	\centering	
	\includegraphics{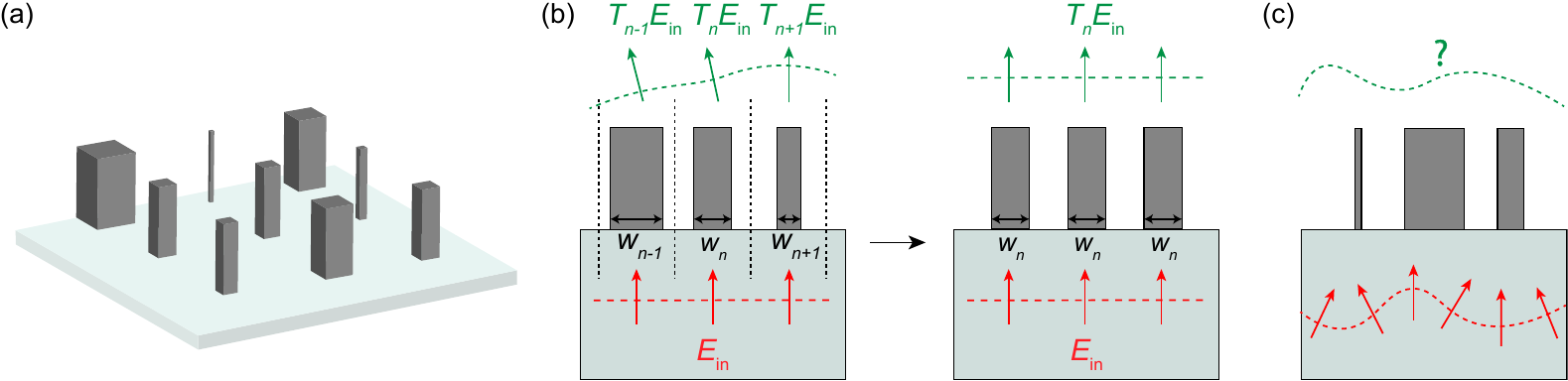}	
	\caption{Conventional technique for modeling metasurfaces. (a) An illustration of a metasurface composed of an aperiodic array of meta-atoms arranged on a periodic lattice. (b) In the conventional modeling technique, the transmitted field at each unit cell is assumed to be proportional to the incident field at the same unit cell. The local transmission coefficient for each meta-atom is obtained by assuming that the structure is locally periodic and is illuminated with a a plane wave. (c) The conventional modeling method is not accurate when the meta-atoms vary rapidly or in cases that the incident wave is not a plane wave.}
	\label{fig:FigOneLabel}
\end{figure}

The conventional approach suffers from several shortcomings and limitations. First, the level of approximation involved in assigning local transmission coefficients to individual unit cells (or meta-atoms) is unclear. As a result, it is unclear if the lattice constant of a metasurface can be considered as the resolution of wavefront shaping, and if it cannot, what is the actual wavefront shaping resolution when using a metasurface? In fact, this method is only valid for slowly varying structures and it is known to lead to inaccurate estimations of efficiencies when the meta-atom geometry changes rapidly from one cell to the next \cite{14-Arbabi2015}. Second, the local transmission coefficients, which are assigned to individual meta-atoms, are obtained for the normal incidence and cannot be used to accurately model the metasurface response to arbitrary incident waves. When the incident wave is a known oblique incident plane wave, this issue can be alleviated by finding the transmission coefficients of the periodic arrays for the known incident angle instead of the normal incidence \cite{18-Arbabi2016,19-Arbabi2017a}, but this workaround is not applicable for more general incident waves (Fig.~\ref{fig:FigOneLabel}c). Third, the conventional metasurface model reduces metasurfaces to simple phase masks (i.e., transparencies) that essentially offer the same features as phase masks implemented using other approaches (e.g., effective medium or kinoform). Compared to other implementations of phase masks, metasurfaces provide several advantages such as simpler fabrication, better performance when implementing rapidly varying phase profiles, and the possibility of controlling chromatic dispersion \cite{20-Arbabi2017b,21-Zhao2015,22-Yuan2017} and polarization \cite{15-Arbabi2015b,23-Kruk2016,24-Pfeiffer2013,25-Yang2014}. Nevertheless, the phase mask model misses a wide range of conceptually interesting and practically important features of metasurfaces that are due to the non-locality of the scattering phenomenon. The non-locality refers to the property that the excitation of a unit cell of the metasurface causes non-zero responses at other unit cells of the metasurface. As we will demonstrate here using an example, the waves transmitted through a metasurface are not local and the phase mask model, which is a local model, cannot be used to exploit the rich opportunities offered by the non-local response of metasurfaces in designing novel optical components. To address this issue, here we introduce the concept of discrete-space impulse response (DSIR) for metasurfaces and employ it to model non-locality in metasurfaces.

The DSIR is a novel technique for modeling linear metasurfaces that overcomes the shortcomings of the conventional approach and offers insights into the interaction of waves with metasurfaces. We introduce the DSIR concept and present a couple of examples of its application that also serve as a confirmation for the validity and accuracy of the DSIR technique. We also introduce two approximations that significantly reduce the computational cost of the DSIR technique while achieving better accuracy than the conventional approach. \\

\section*{Discrete-space impulse response concept}
\begin{figure}[]
  \centering
  \includegraphics[]{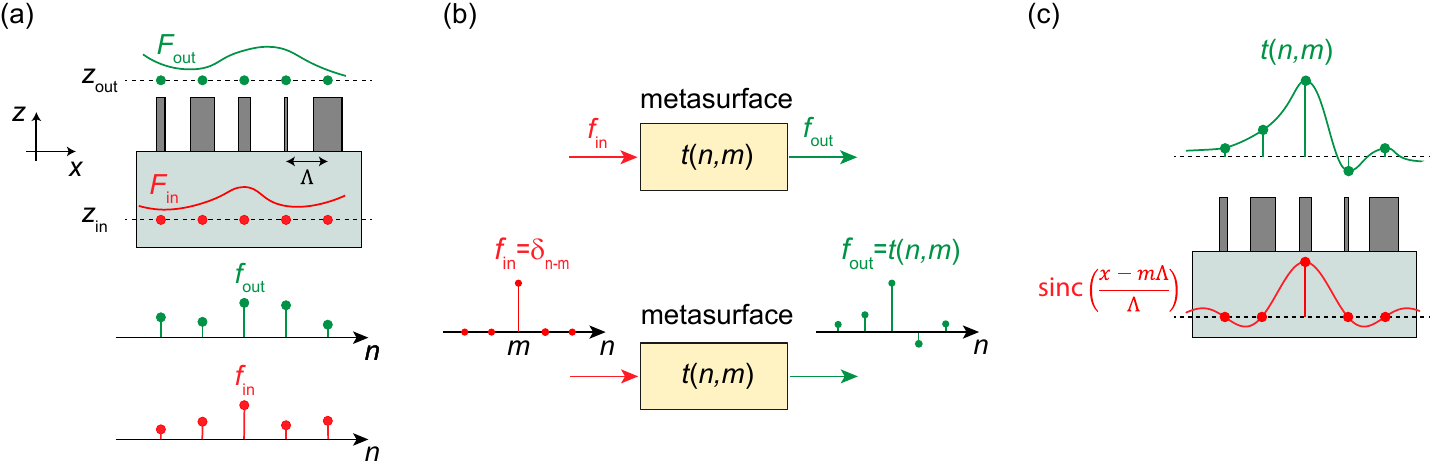} 
  \caption{DSIR concept. (a) Illustration of a metasurface illuminated with an incident wave with the field $F_{\mathrm{in}}$. Incident and outgoing waves on the input and output reference planes, which are at $z_{\mathrm{in}}$ and $z_{\mathrm{out}}$, respectively, are sampled at the lattice sites of the metasurface. The sampled incident and outgoing fields can be considered as discrete signals $f_{\mathrm{in}}(n)$ and $f_{\mathrm{out}}(n)$. (b) A metasurface may be considered as a discrete linear system with an impulse response $t(n,m)$ that maps $f_{\mathrm{in}}(n)$ to $f_{\mathrm{out}}(n)$. The impulse response is the system’s output to a discrete delta input. (c) The procedure for obtaining the impulse response of a metasurface. The metasurface is excited with a reconstruction function $W(x)$, which is shown as a sinc function, that is centered at the $m^\mathrm{th}$ unit cell of the metasurface and $t(n,m)$ is obtained by sampling the outgoing field at the $n^\mathrm{th}$ metasurface unit cell.}
  \label{fig:FigTwoLabel}
\end{figure}

In this section, we explain DSIR concept for modeling linear and non-diffractive metasurfaces with periodic lattices. For simplicity, we explain the concept using a 1D transmissive metasurface but the idea and approach are general and can be readily extended for modeling more general metasurfaces (i.e., 2D or reflective metasurfaces). Consider the 1D metasurface schematically shown in Fig.~\ref{fig:FigTwoLabel}a. The metasurface is composed of an array of potentially different meta-atoms that are positioned at the lattice sites of a periodic lattice with the lattice constant of $\mathrm{\Lambda}$. The metasurface is surrounded by materials with refractive indices of $n_1$ and $n_2$ at the bottom and on the top, respectively, which can represent the cladding and the substrate. 

Assume that an incident wave with an arbitrary wavefront is propagating in the material with a refractive index of $n_1$ and is incident on this metasurface from the bottom side. We choose a plane parallel to the plane of the metasurface at $z=z_{\mathrm{in}}$ and refer to it as the “input reference plane”. The input reference plane can be chosen arbitrarily. According to the surface equivalence theorem \cite{26-harrington1961time}, the tangential components of either the electric field or the magnetic field of the incident wave on the input reference plane uniquely characterize the incident wave.    
Let $F$ depict the transverse components (with respect to $z$) of either the electric or the magnetic field. For example, for the 1D metasurface shown in Fig.~\ref{fig:FigTwoLabel}a, $F$ may depict $E_y$ for transverse electric waves and $H_y$ for transverse magnetic waves. The spatial Fourier transform of the field of the incident wave ($F_{\mathrm{in}}$) on the input reference plan is given by
\begin{equation}
\widetilde{F}_{\mathrm{in}}(k_x,z_{\mathrm{in}})=\int_{-\infty}^{\infty} \! F_{\mathrm{in}}(x,z_{\mathrm{in}})\mathrm{e}^{jk_xx} \, \mathrm{d}x.
\end{equation}
If the incident wave is emitted by a source that is at least a few wavelengths away from the metasurface, which is a valid assumption in many practical cases, then the wave on the input reference plane is composed of only traveling waves and has no evanescent components. As a result, the transverse components of the electric (or magnetic) field of the incident wave on the input reference plane (i.e., ${F_{\mathrm{in}}}$) are bandlimited in the spatial Fourier plane of $k_x$. In other words, $\widetilde{F}_{\mathrm{in}}(k_x,z_{\mathrm{in}})$ is zero for $|k_x|>n_1k_0$. The limited spatial bandwidth of the incident field is because it is a propagating wave and can be considered as a superposition of fields of propagating plane waves (no contributions from evanescent plane waves exist in its expansion). The maximum spatial bandwidth of $F_{\mathrm{in}}(x)$ is $n_1k_0$; therefore, it can be uniquely reconstructed from its samples if it is sampled at least at the Nyquist rate with a maximum sampling period of $\Delta x={\frac{2\pi}{2n_1k_0}}=\lambda_1/2$, where $\lambda_1$ is the wavelength of the incident wave in the substrate. Because the metasurface lattice is non-diffractive in the substrate $\mathrm{\Lambda}<\lambda_1/2$ we can use the lattice sites as the sampling locations. Therefore, the incident wave on the input reference plane can be represented by its samples at the lattice sites of the metasurface.

Similar to the input reference plane, we can choose an “output reference plane” above the metasurface plane at $z=z_{\mathrm{out}}$ (as schematically shown in Fig.~\ref{fig:FigTwoLabel}a). For example, the output reference plane can be selected slightly above the plane right on top of the meta-atoms. The field of the transmitted waves on the output reference plane $F_{\mathrm{t}}(x)=F(x,z_{\mathrm{out}})$ might not be bandlimited to $n_2k_0$ if the output reference plane is selected close to the meta-atoms where the evanescent waves have not yet decayed significantly. However, the frequency components of $F_{\mathrm{t}}(x)$ with $|k_x|>n_2k_0$ do not propagate away from the metasurface and do not contribute to the transmitted waves measured at any locations which are at least a few wavelengths away from the metasurface plane. As a result, we can sample $F_{\mathrm{t}}(x)$ at a high enough sampling rate to avoid aliasing, and then filter it using an ideal low-pass filter which is equal to 1 for spatial frequencies of $|k_x|\leq n_2k_0$, and 0 otherwise. The result of this filtering procedure (which we refer to as the outgoing wave $F_{\mathrm{out}}(x)$) has the same propagating spatial components as $F_{\mathrm{t}}(x)$, thus it leads to the same transmitted wave after propagating a few wavelengths away from the metasurface. Because $F_{\mathrm{out}}(x)$ is bandlimited to $n_2k_0$ it can be sampled with the period of $\mathrm{\Lambda}<\lambda_2/2$. Also, the knowledge of the tangential components of $F$ on any plane above the metasurface is sufficient to determine the transmitted fields uniquely in the region above that plane. As a result, the samples of $F_{\mathrm{out}}(x)$ obtained at the metasurface lattice sites will completely characterize the transmitted wave at any location that is at least a few wavelengths away from the metasurface plane.

The sampled input and output fields are schematically shown in Fig.~\ref{fig:FigTwoLabel}a as discrete signals $f_{\mathrm{in}}(n)=F_{\mathrm{in}}(n\mathrm{\Lambda})$ and $f_{\mathrm{out}}(n)=F_{\mathrm{out}}(n\mathrm{\Lambda})$ (i.e., sequences of complex numbers, $n$: integer). As Fig.~\ref{fig:FigTwoLabel}b shows, the metasurface transforms input discrete signals to output discrete signals. For linear metasurfaces (i.e., at low incident power levels where nonlinear effects are negligible), the transformation is linear and we can define its discrete-space impulse response (DSIR) $t(n,m)$ as the output signal $f_{\mathrm{out}}(n)$ corresponding to $f_{\mathrm{in}}(n)=\delta_{n-m}$, where $\delta_i$ is the Kronecker delta function that is equal to 1 for $i=0$ and zero otherwise. (Fig.~\ref{fig:FigTwoLabel}b).

According to the Nyquist sampling theorem, the continuous fields $F_{\mathrm{in/out}}(x)$ can be exactly reconstructed from their sampled values (i.e., the discrete signals) by interpolating the discrete signals using a specific interpolation function $W(x)$ that is the impulse response of a low-pass reconstruction filter. In the 1D case, the interpolation function $W$ can be selected as a sinc function ($W(x)=\mathrm{sinc}(x/\mathrm{\Lambda})=\mathrm{sin}(\pi x/\mathrm{\Lambda})/(\pi x /\mathrm{\Lambda})$) and the fields can be written as \cite{27-mersereau1984multidimensional}
\begin{equation}
	F_{\mathrm{in/out}}(x)=\sum_{n=-\infty}^{\infty}f_{\mathrm{in/out}}(n)W(x-n\Lambda)=\sum_{n=-\infty}^{\infty}f_{\mathrm{in/out}}(n)\mathrm{sinc}(\frac{x-n\mathrm{\Lambda}}{\mathrm{\Lambda}}).
\end{equation}
According to (2), the impulse response $t(n,m)$ can be interpreted as the samples of the outgoing field $F_{\mathrm{out}}(x)$ when the incident field is $F_{\mathrm{in}}(x)=W(x-m\mathrm{\Lambda})$. For the selection of $W(x)=\mathrm{sinc}(x/\mathrm{\Lambda})$, the incidence field $F_{\mathrm{in}}=\mathrm{sinc}(\frac{x}{\mathrm{\Lambda}}-m)$ is the field of a wave focused with a numerical aperture of $({\lambda}_1/2)/\mathrm{\Lambda}$ at the location of the $m^\mathrm{th}$ meta-atom. This is schematically shown in Fig.~\ref{fig:FigTwoLabel}c. Also, for this selection, the incident wave amplitude for the impulse response ($F_{\mathrm{in}}(x)$) passes through zero at all lattice sites except for the $m^\mathrm{th}$ one and decreases away from the $m^\mathrm{th}$ lattice site. As a result, the main contributions to $t(n,m)$ are expected to be from the $m^\mathrm{th}$ meta-atom and its close neighbors. For local metasurfaces, we expect impulse responses similar to the one shown in Fig.~\ref{fig:FigTwoLabel}c where $|t(n,m)|$ decreases with increasing $|n-m|$.
The discrete-space impulse response $t(n,m)$ completely characterizes the metasurface response. The samples of the outgoing wave for any arbitrary incident wave with a field $F_{\mathrm{in}}(x)$ can be obtained according to
\begin{equation}
	f_{\mathrm{out}}(n)=\sum_{m=-\infty}^{\infty}f_{\mathrm{in}}(m)t(n,m),
\end{equation}
where $f_{\mathrm{in}}(n)=F_{\mathrm{in}}(n\mathrm{\Lambda})$, and $F_{\mathrm{out}}(x)$ can be reconstructed from $f_{\mathrm{out}}(n)$ according to (2). Equation (3) can be written in a matrix form as
\begin{equation}
	f_{\mathrm{out}}=\boldsymbol{t} f_{\mathrm{in}}
\end{equation}
where $f_{\mathrm{in}}$ and $f_{\mathrm{out}}$ are vectors whose $n^\mathrm{th}$ elements are equal to $f_{\mathrm{in}}(n)$ and $f_{\mathrm{out}}(n)$, respectively, and $\boldsymbol{t}$ is a matrix with elements $t_{nm}=t(n,m$). The assumption of locality used in the conventional metasurface analysis and design is equivalent to assuming that $\boldsymbol{t}$ is diagonal. As we will show through an example in the next section, the off-diagonal elements of $\boldsymbol{t}$ are nonzero for typical metasurfaces. For metasurfaces with small coupling between meta-atoms, the amplitudes of the off-diagonal elements should decrease as their distances to the diagonal increase. 

The metasurface matrix $\boldsymbol{t}$ can be expressed in any basis, but the basis used in the DSIR approach (i.e., $W(x-m\mathrm{\Lambda})$) leads to a compact and intuitive representation of wave transformations offered by local metasurfaces. For example, the fields on the input and output reference planes ($F_{\mathrm{in}}$ and $F_{\mathrm{out}}$) may be expanded in terms of a continuous basis of plane waves propagating along different directions. However, the matrix $\boldsymbol{t}$ of a typical aperiodic metasurface (e.g., a metalens) in the plane wave basis is not sparse because the metasurface transforms an incident plane wave into a continuous superposition of outgoing plane waves. This approach can be applied to any linear metasurface, irrespective of the physics governing the interactions of light with the metasurface. Hence, as long as the simulations employed for finding DSIR responses properly model the light interactions with meta-atoms, the method is valid. Also, no approximation is involved in the derivation of (3) and transmitted waves reconstructed from $f_{\mathrm{out}}$ are accurate in regions that are at least a few wavelengths away from the metasurface where the evanescent waves have significantly decayed. In practice, the impulse responses should be truncated which introduces an error that can be adjusted by selecting the truncation range.

\section*{Computation and applications of DSIR}
In this section, we describe the computation procedure of DSIR using a couple of examples and illustrate its applications in determining the response of metasurfaces to arbitrary incident waves. For these demonstrations, we use the simple 1D transmissive metasurface platform shown in Fig.~\ref{fig:FigThreeLabel}a. The metasurface operates at $\lambda=\mathrm{1}$ $\upmu$m and is composed of 0.8-$\upmu$m-tall rectangular silicon bars with a refractive index of 3.46 that are arranged on a periodic lattice with a lattice constant of $\mathrm{\Lambda}=\mathrm{0.5}$ $\upmu$m. The substrate and top cladding are assumed to be air. The transmission of periodic metasurfaces implemented using this platform for a normally-incident TM-polarized plane wave (with magnetic field along the $y$ direction) is also shown in Fig.~\ref{fig:FigThreeLabel}a and is used for designing different components discussed in this and the next sections.

\subsection*{DSIR of non-diffractive periodic metasurfaces}
\begin{figure}[t]
  \centering
  \includegraphics[]{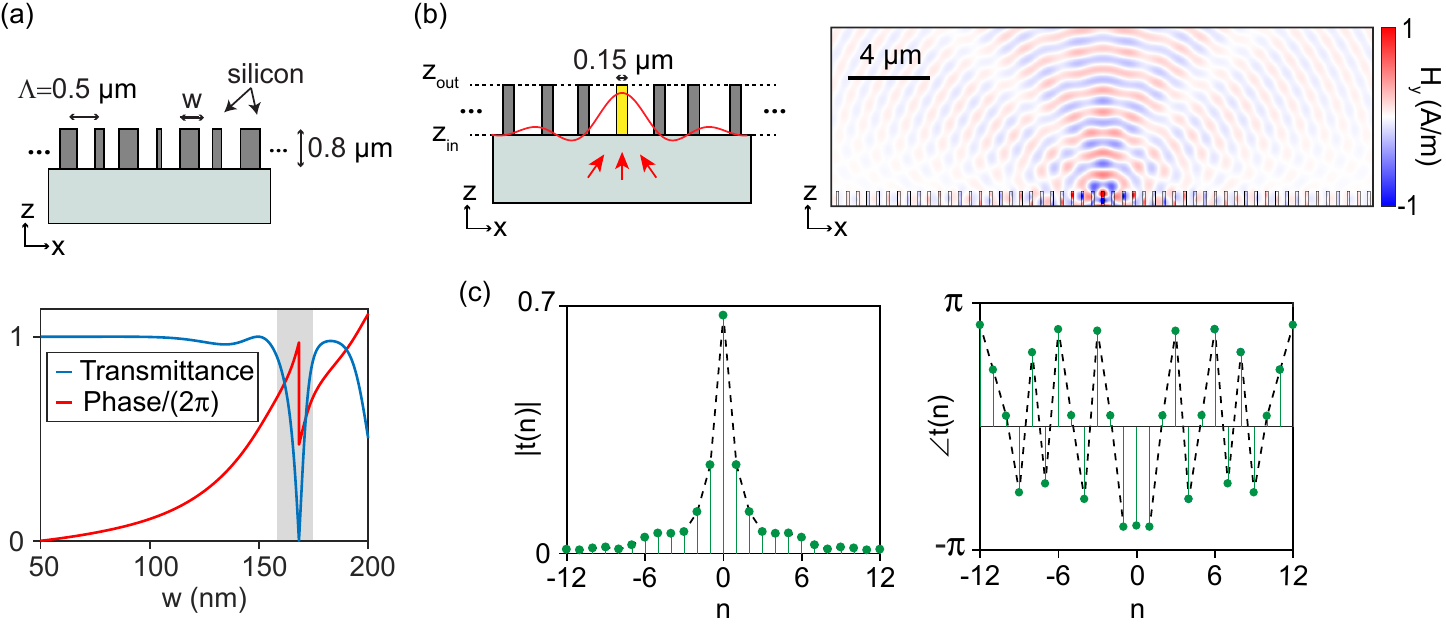} 
  \caption{DSIR of a periodic metasurface. (a) Schematic illustration of a 1D metasurface used for the demonstration of DSIR computation and applications. The metasurface is composed of silicon bars with different width ($w$) and its transmission as a function of $w$ is shown. (b) Schematic of an infinite periodic metasurface based on the design shown in (a). Input and output reference planes are selected at the bottom and top of the meta-atoms, respectively. The metasurface is excited by an incident field whose magnetic field at the input reference plane is given by $\vec{H}=\mathrm{sinc}(\frac{x}{\mathrm{\Lambda}})\hat{y}$, and a snapshot of the transmitted magnetic field is shown on the right. (c) Amplitude and phase of the DSIR of the metasurface shown in (b). n represents the discrete spatial coordinate.}
  \label{fig:FigThreeLabel}
\end{figure}

Consider the periodic metasurface shown in Fig.~\ref{fig:FigThreeLabel}b where all the meta-atoms have equal width of $w=\mathrm{150}$ nm and the input and output reference planes are selected at the bottom and top of the meta-atoms, respectively. Because of the periodicity, the DSIR of such a periodic metasurface $t(n,m)$ is only a function of the difference between $n$ and $m$, thus it suffices to find $t(n)=t(n,0)$ and DSIR can be computed using a single simulation. As we discussed in the previous section, the DSIR is determined by finding samples of the outgoing field for the incident field $F_{\mathrm{in}}(x)=\mathrm{sinc}(\frac{x}{\mathrm{\Lambda}}-m)=\mathrm{sinc}(\frac{x}{\mathrm{\Lambda}})$. Therefore, to obtain $t(n)$ for the periodic metasurface shown in Fig.~\ref{fig:FigThreeLabel}b, the metasurface was illuminated by a TM-polarized wave propagating along the $+z$ direction whose magnetic field at the input reference plane is given by $\vec{H}=\mathrm{sinc}(\frac{x}{\mathrm{\Lambda}})\hat{y}$. The incident wave was generated by an electric surface current density $\vec{J_\mathrm{s}}=-2\mathrm{sinc}(\frac{x}{\mathrm{\Lambda}})\hat{x}$ placed on the input reference plane.

The response of the metasurface to $F_{\mathrm{in}}$ was found using a finite element method simulation using a commercial software and a snapshot of the magnetic field is shown in Fig.~\ref{fig:FigThreeLabel}b. The transmitted fields on the output reference plane have significant evanescent components because the plane is chosen close to the meta-atoms. To reduce numerical errors and the required sampling frequency, the transmitted field was sampled at a plane 0.5 $\upmu$m above the output reference plane, low-pass filtered to remove any remaining evanescent components, backpropagated to the $z=z_{\mathrm{out}}$ plane, and was sampled at $x=n\mathrm{\Lambda}$. For forward and backward propagation through homogenous materials, we used the plane wave expansion technique \cite{28-born2013principles}. The modulus and phase of DSIR of the periodic metasurface (Fig.~\ref{fig:FigThreeLabel}b) were obtained using this procedure and are shown in Fig.~\ref{fig:FigThreeLabel}c. The DSIR data shown in Fig.~\ref{fig:FigThreeLabel}c completely characterizes the periodic metasurface and the outgoing waves for any arbitrary incident waves can be obtained from (3) which reduces to a discrete convolution operation for non-diffractive periodic metasurfaces. The DSIR shown in Fig.~\ref{fig:FigThreeLabel}c is localized around $x=0$, but has a finite width and is not a discrete delta function. Therefore, while the interaction of the metasurface with an incident wave can be considered relatively local, the interaction length is a few unit cells and the conventional technique, which assumes interactions are limited to a single unit cell, does not accurately model wave interactions with metasurfaces.

The DSIR of a non-diffractive periodic metasurface can also be obtained from its transmission coefficients for plane waves incident on the metasurface at different angles. As it is shown in Supplementary Note 1, the complex-valued transmission coefficient of the metasurface $T(\theta)$ for a plane wave incident at an angle $\theta$ is equal to $\widetilde{t}(2\pi \mathrm{sin}(\theta)\mathrm{\Lambda} / \lambda_1)$ where $\widetilde{t}(\omega)$ is the Fourier transform of $t(n)$. In other words, the DSIR of a non-diffractive periodic metasurface is the inverse Fourier transform of the angular spectrum of its transmission coefficient. We verified this relation for the metasurface shown in Fig.~\ref{fig:FigThreeLabel}b by finding its transmission coefficient for plane waves incident at different angles, taking its Fourier transform, and comparing the result with the DSIR shown in Fig.~\ref{fig:FigThreeLabel}c which was obtained using a finite element simulation.

The Fourier transform relation between DSIR and transmission coefficient offers several benefits. First, the transmission coefficient of plane waves incident on periodic structures can be computed using periodic boundary conditions or Fourier modal techniques such as the rigorous coupled-wave analysis (RCWA) technique, thus reducing the DSIR computational cost. Second, the Fourier transform provides an intuitive relation between the transmission coefficient and DSIR. For example, the transmission coefficient of an ideal local metasurface, whose DSIR is a discrete delta function, should be independent of the incident angle, or a resonant feature in the angular spectrum of a metasurface indicates a slow decay of its DSIR. Third, it clarifies the relation of the DSIR and the approximation used in the conventional technique. The normal-incidence transmission coefficient that is used in the conventional method is equal to $\widetilde{t}(0)$, which via the Fourier transform relation, is equal to the sum of all DSIR elements $T(0)=\sum_{n}t(n)$. Fourth, as we have shown using the Fourier transform relation in Supplementary Note 2, the Euclidean norm of the DSIR is bounded by 1 (i.e., $\sum_{n}|t(n)|^2 \leq 1$), hence the DSIR has a finite width. 

\subsection*{Analysis of metasurface beam deflectors using the DSIR technique}
\begin{figure}[]
  \centering
  \includegraphics[width=1.0\textwidth]{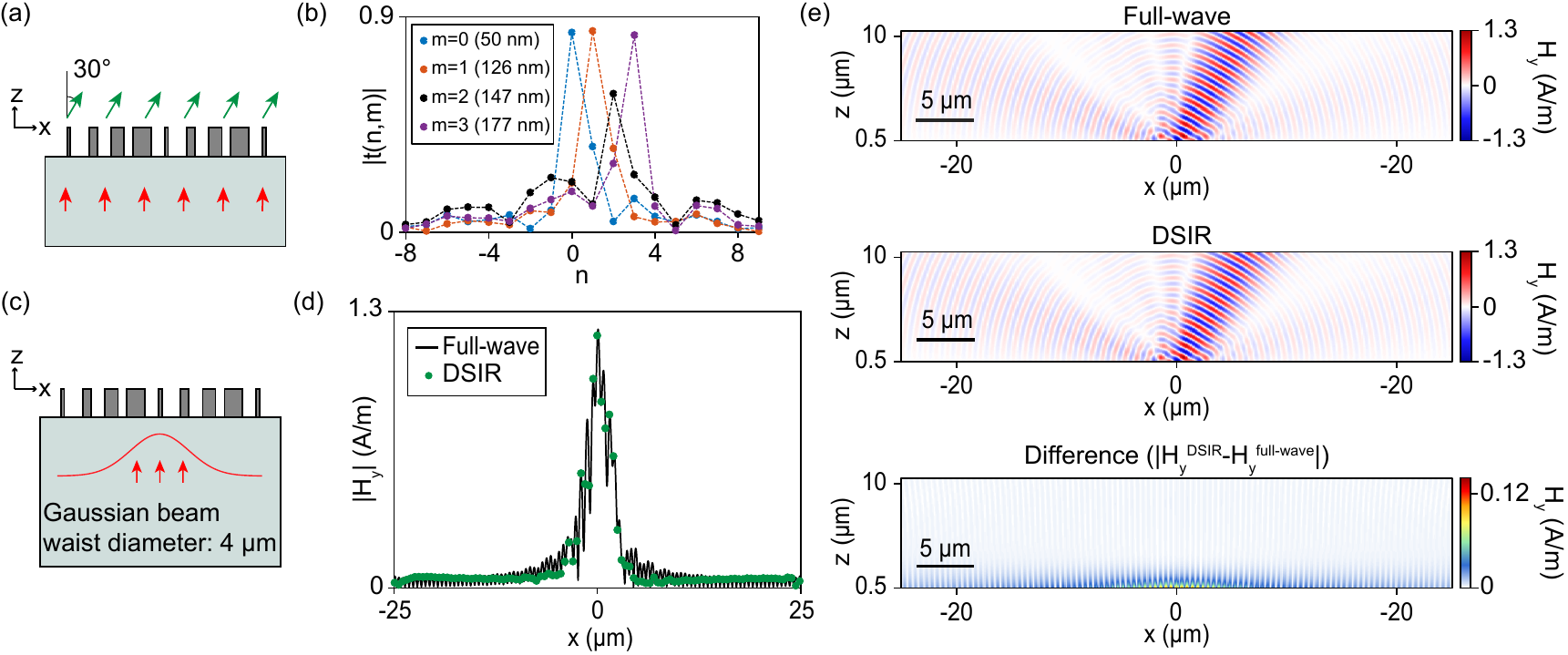} 
  \caption{DSIR of a metasurface beam deflector. (a) Schematic of a 30$^\circ$ metasurface beam deflector. The beam deflector is implemented using the platform shown in Fig. 3a and is composed of meta-atoms with widths of 50 nm, 126 nm, 147 nm, and 177 nm. (b) Modulus of DSIR for incident waves centered at the four meta-atoms. The meta-atom’s width at the excitation center for different values of $m$ are listed in the legend. (c) Schematic of the beam deflector illuminated by a Gaussian beam with a waist diameter of 4 $\upmu$m. (d) Modulus of the magnetic field of the outgoing wave on a plane 0.5 $\upmu$m above the top of the meta-atoms for the metasurface shown in (c). The results obtained using full-wave simulation and the DSIR technique are shown. (e) Snapshots of the magnetic field in $z>$0.5 $\upmu$m computed using a full-wave simulation and the DSIR technique and their difference.}
  \label{fig:FigFourLabel}
\end{figure}

To demonstrate the accuracy of the DSIR technique and to illustrate its application in full characterization of a more general metasurface, we consider a wave interaction with a metasurface beam deflector. Consider the beam deflector schematically shown in Fig.~\ref{fig:FigFourLabel}a that is designed using the metasurface design presented in Fig.~\ref{fig:FigThreeLabel}a. The metasurface deflects normally incident TM-polarized plane waves by 30$^\circ$ and is composed of four different meta-atoms with widths of 50 nm, 126 nm, 147 nm, and 177 nm, corresponding to relative phase shifts of 0, $\pi/$2, $\pi$, and 3$\pi/$2, respectively. Because the beam deflector is periodic with a period of 4$\mathrm{\Lambda}$, its DSIR (i.e., $t(n,m)$) can be computed using four simulations with incident fields centered at four different unit cells ($F_{\mathrm{in}}=\mathrm{sinc}(x/\mathrm{\Lambda}-m)$ for $m=0,1,2,3$). A larger number of simulations are required for determining the DSIR of general aperiodic metasurfaces, and we discuss techniques for reducing the DSIR computational cost in the next section; however, the four meta-atom beam deflector serves as a proof-of-concept example for demonstrating the DSIR technique. We chose the input reference plane at the bottom of the meta-atoms at $z_{\mathrm{in}}=-$0.8 $\upmu$m and the output reference plane at $z_{\mathrm{out}}=\mathrm{0}$ which is the top surface of the meta-atoms. The DSIR was computed by finite element simulations using a procedure similar to the one detailed in the previous section, and the results are presented in Fig.~\ref{fig:FigFourLabel}b. As this figure demonstrates, although the responses are mostly localized around the excited meta-atoms, they have significant values at other adjacent unit cells of the metasurface. Such non-local effects were ignored in the conventional method and caused inaccurate results.

By shifting the computed DSIRs to the location of similar meta-atoms of the beam deflector, we can form the matrix $\boldsymbol{t}$ and find the outgoing waves for any arbitrary incident waves. As an example, we assumed the beam deflector is illuminated by a normally-incident TM-polarized Gaussian beam with a beam diameter of 4 $\upmu$m whose waist is at the $z_{\mathrm{in}}$ plane (Fig.~\ref{fig:FigFourLabel}c). Figure~\ref{fig:FigFourLabel}d shows the outgoing magnetic field (i.e., the propagating component of the transmitted field) at the output reference plane obtained using the DSIR technique (Eq. (4)) and a full-wave simulation with the Gaussian incident beam. The two results shown in Fig.~\ref{fig:FigFourLabel}d are in good agreement, thus confirming the accuracy of the DSIR technique. Figure~\ref{fig:FigFourLabel}e shows snapshots of the magnetic field in the region above the meta-atoms computed using a full-wave simulation and by propagating the DSIR field on the output reference plane using the plane wave expansion method. The difference between the two fields is also shown in Fig.~\ref{fig:FigFourLabel}e, and represents the evanescent fields that are large only close to the metasurface.

The DSIR technique accurately predicts metasurface response to any arbitrary incident field with no approximations involved. This technique finds the actual response of each different meta-atom in a metasurface and provides a matrix that completely characterizes the whole structure. However, for more general metasurfaces where all meta-atoms are different, the computation of DSIR involves a large number of simulations. To reduce the complexity of the computations, we introduce two approximation methods to this technique. For metasurfaces with slowly varying meta-atoms, we can approximate each meta-atom to be located in a periodic structure with the same periodicity (local periodicity approximation (LPA)). Also, as the amplitude of the DSIR excitation (i.e., the sinc function) drops away from the excitation’s center, the DSIR is expected to be localized and only depend on the meta-atom at the excitation center and a few of its neighbors. The localized feature of the DSIR allows for its estimation from the knowledge of the meta-atoms in a neighborhood of the excitation center and introduces a pathway for improving the accuracy of the conventional technique. In the following section, we discuss two methods for the estimation of DSIR using a limited number of meta-atoms that reduce its computational cost significantly.

\section*{Approximate techniques for computing DSIR}
\subsection*{Local periodicity approximation of slowly varying metasurfaces}
Meta-atom geometries vary slowly in metasurfaces such as beam deflectors with small deflection angles and low numerical aperture metalenses. In such metasurfaces, meta-atoms are approximately the same in the DSIR excitation region, thus the metasurface can be locally approximated by a periodic metasurface. Using this approximation, DSIRs are assigned to meta-atoms instead of unit cells, thus reducing the number of simulations required for a general metasurface to the number of distinct meta-atoms. The conventional technique also uses the local periodicity approximation; however, in contrast to a single transmission coefficient that is used in the conventional approach, the DSIR contains incident-angle-dependent information of the meta-atom and leads to more accurate results for incident waves with arbitrary wavefronts. As discussed in the previous section, the DSIR of non-diffractive periodic structures can be obtained from the angular spectrum of their transmission coefficients which by itself can be computed at a low computational cost using a Fourier modal method \cite{29-moharam1981rigorous}.

\begin{figure}[]
  \centering
  \includegraphics[width=1.0\textwidth]{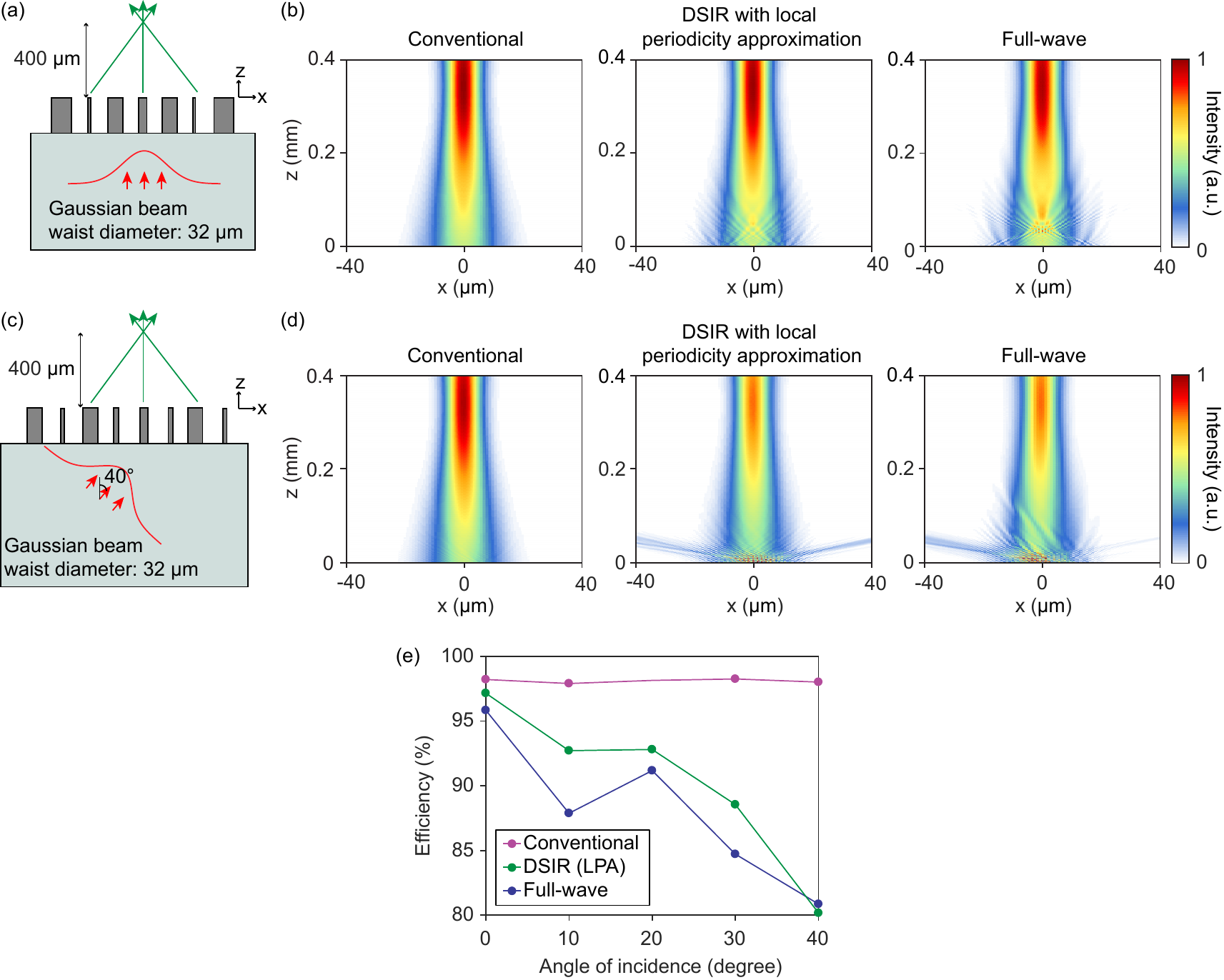} 
  \caption{DSIR with the local periodicity approximation. (a) Schematic of a low-numerical-aperture metalens with a focal length of 400 $\upmu$m focusing a normally incident Gaussian beam with a waist diameter of 32 $\upmu$m. (b) Intensity distributions of the light transmitted through the metalens shown in (a) computed using the conventional technique, DSIR with local periodicity approximation, and a full-wave simulation. (c) Schematic of an off-axis metalens with a focal length of 400 $\upmu$m that is designed to focus light incident at 40$^\circ$. The metalens is illuminated with a Gaussian beam with a waist diameter of 32 $\upmu$m that is incident at 40$^\circ$. (d) Intensity distributions of the light transmitted through the metalens shown in \textbf{c} computed using the conventional technique, DSIR with local periodicity approximation, and a full-wave simulation. (e) Estimated focusing efficiencies of five different metalenses designed for different incident angles (0$^\circ$ to 40$^\circ$, 10$^\circ$ steps) obtained using the conventional and DSIR with local periodicity approximations.  Accurate focusing efficiency values computed using full-wave simulations are also shown. The metalenses are similar to the ones shown in (a) and (c) and are illuminated by a Gaussian beam with a waist diameter of 32 $\upmu$m that is incident at the metalens's design angle.}
  \label{fig:FigFiveLabel}
\end{figure}

To evaluate the performance of this approximation, we designed a number of low numerical aperture metalenses that focus light incident at different angles to a point 400 $\upmu$m away from their surfaces. Figure~\ref{fig:FigFiveLabel}a shows one of the metalenses designed for normal incidence that is illuminated by a normally incident TM-polarized Gaussian beam with a waist diameter of 32 $\upmu$m. 
The intensity distributions of light transmitted through this metalens were obtained using the conventional technique, the DSIR technique with the local periodicity approximation, and a full-wave simulation are presented in Fig.~\ref{fig:FigFiveLabel}b. As Fig.~\ref{fig:FigFiveLabel}b shows, all three approaches predict similar focal point intensities. Figure~\ref{fig:FigFiveLabel}c displays another metalens that is designed for an incident angle of 40$^\circ$ and is illuminated with a similar Gaussian beam incident at 40$^\circ$. Figure~\ref{fig:FigFiveLabel}d shows the intensity distributions of the transmitted light of this metalens computed using different approaches. As the results presented in Fig.~\ref{fig:FigFiveLabel}d show, the DSIR with local periodicity approximation provides a more accurate estimation of the focused light’s intensity than the conventional approach.

Similar metalenses with incident angles of 10$^\circ$, 20$^\circ$, and 30$^\circ$ were also designed and simulated and their focusing efficiencies computed using the three different approaches are shown in Fig.~\ref{fig:FigFiveLabel}e. The focusing efficiency represents the fraction of transmitted light that is focused to a 10-$\upmu$m-wide area around the focal point when the metalens is illuminated by a Gaussian beam with a waist diameter of 32 $\upmu$m that is incident at the metalens design angle. As the results shown in Fig.~\ref{fig:FigFiveLabel}e indicate, the local periodicity approximation of DSIR leads to more accurate estimations of the focusing efficiency of metalenses than the conventional modeling approach. The result is expected because the DSIR with local periodicity approximation reduces to the conventional technique by further approximating the DSIR by a discrete delta function equal to the sum of its elements.

\subsection*{Locality approximation of DSIR}
\begin{figure}[]
  \centering
  \includegraphics{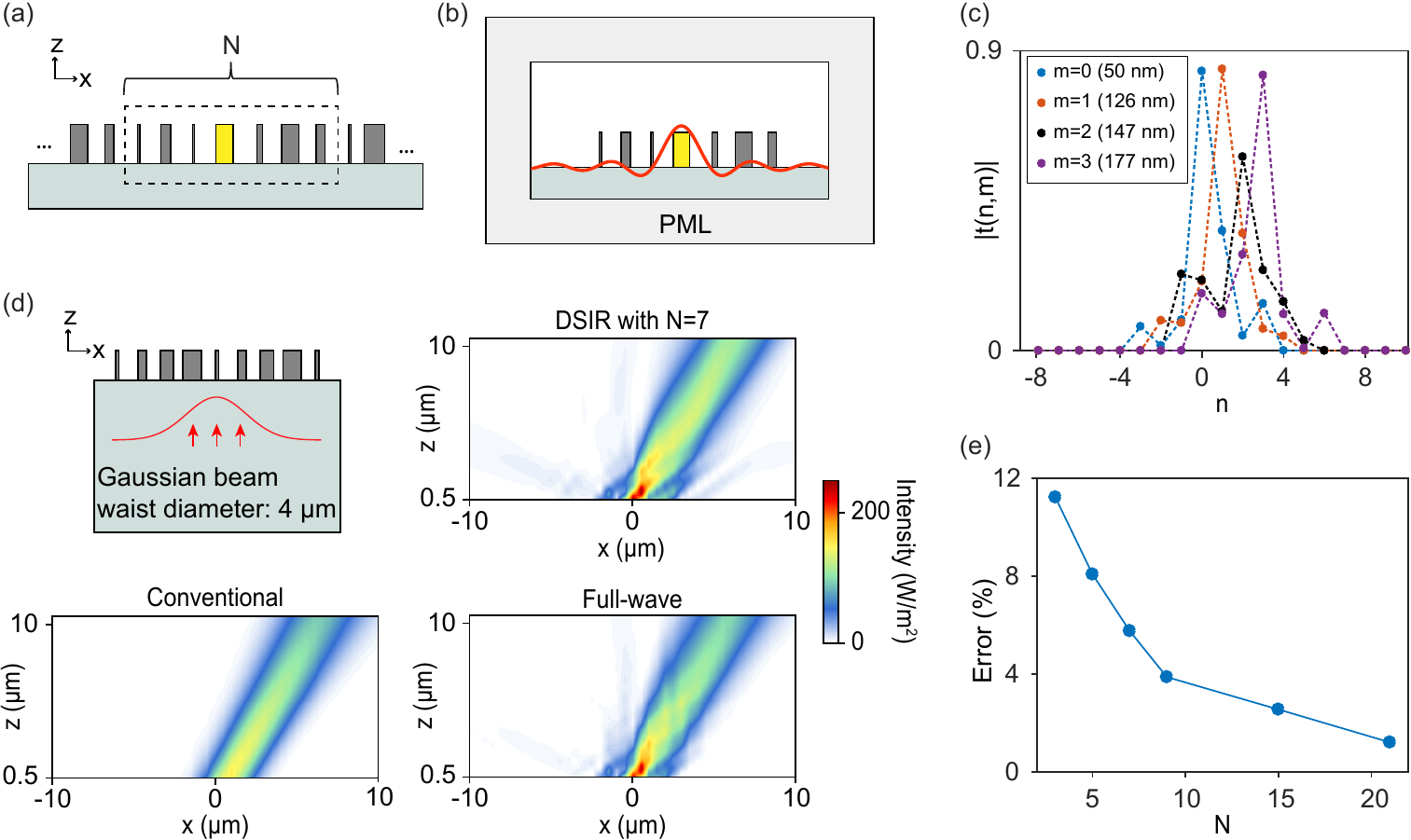} 
  \caption{Locality approximation of DSIR. (a) Illustration of the windowing idea used in the locality approximation. A finite number of meta-atoms ($N$) close to the excitation center are considered for the DSIR computation. The meta-atom at the excitation center is shown by a different color. (b) Schematic of the structure simulated for the DSIR estimation using the locality approximation. The simulation domain is surrounded by a perfectly matched layer (PML) to prevent reflections from its boundaries. (c) Modulus of the approximate DSIR of the beam deflector shown in Fig. 4 obtained using the locality approximation with $N=7$. The width of the meta-atom at the excitation center for different values of $m$ are listed in the legend. (d) Schematic and the intensity distribution of the transmitted light for the beam deflector of Fig. 4 whose approximate DSIR is shown in (c). The beam deflector is illuminated by a normally incident Gaussian beam with a beam diameter of 4 $\upmu$m, and the intensity distribution computed using the DSIR with the locality approximation and $N=7$ is shown. Intensity distributions obtained using the conventional approach and a full-wave simulation are also shown for comparison. (e) Error of the locality approximation as a function of the number of meta-atoms in each window for the beam deflector shown in (d).}
  \label{fig:FigSixLabel}
\end{figure}

The periodic approximation provides an accurate result for slowly varying metasurfaces, but its accuracy is reduced for metasurfaces with rapid variations. Direct calculation of the DSIR by exciting a metasurface with sinc functions centered at different unit cells can be used but this approach involves a large number of simulations of the entire metasurface. The excitation amplitude (i.e., the sinc function) is only significant over a finite number of unit cells close the excitation center, thus we can expect that the meta-atoms faraway from the excitation center to have a negligible effect on the metasurface response and its DSIR. By ignoring the effect of faraway meta-atoms, which we refer to as the locality approximation, we can find an approximate DSIR of a metasurface. As it is shown schematically in Fig.~\ref{fig:FigSixLabel}a, only a finite number of meta-atoms ($N$) close to the excitation center are considered (i.e., windowing the metasurface structure).

Windowing reduces the computational cost of each simulation and the number of simulations required for computing the DSIR matrix. An illustration of a simulation for computing DSIR with $N=7$ is shown in Fig.~\ref{fig:FigSixLabel}b. The small size of the simulation domain significantly reduces the simulation’s computational cost. Furthermore, most metasurfaces are composed of a limited number of different meta-atoms, hence a lookup table for the DSIR can be constructed allowing for the computation of the response of any metasurface to any arbitrary incident waves using a single matrix multiplication (i.e., using Eq. (4)).

To study the performance of the locality approximation, we considered the beam deflector shown in Fig.~\ref{fig:FigFourLabel}. The DSIR for this beam deflector obtained using the locality approximation with $N=7$ is shown in Fig.~\ref{fig:FigSixLabel}c, which is comparable with the exact results shown in Fig.~\ref{fig:FigFourLabel}b.

The beam deflector’s response to a  normally incident Gaussian beam was computed using the approximate DSIR shown in Fig.~\ref{fig:FigSixLabel}c, and the result is shown in Fig.~\ref{fig:FigSixLabel}d along with the results obtained using the conventional approach and a full-wave simulation. As the intensity distributions in Fig.~\ref{fig:FigSixLabel}d show, the DSIR with locality approximation is more accurate than the conventional approach. To study the effect of changing the number of meta-atoms in each window and quantify the locality approximation's accuracy, we compared the estimated magnetic field of the outgoing field on a plane 0.5$\upmu$m above the meta-atoms with the full-wave result. We defined the error as the square of the ratio of the norm of the difference between the estimate and the full-wave fields to the norm of the full-wave field. Figure~\ref{fig:FigSixLabel}e shows the error of the locality approximation of DSIR as a function of the number of the meta-atoms in each window ($N$) for the beam deflector. As expected, by increasing $N$ the error of the approximation decreases and the results converge to the full-wave simulation results (Fig.~\ref{fig:FigSixLabel}e). For comparison, the conventional approach’s error for this structure is \texttildelow 30$\%$. As the results shown in Fig.~\ref{fig:FigSixLabel}e display, considering the effect of the nearest neighbors (i.e., $N=3$) reduces the error to less than 12$\%$.

\section*{Development of the DSIR concept for 2D and diffractive metasurfaces}
Although we described the concept of DSIR using 1D metasurfaces, the DSIR concept can be readily extended to 2D metasurfaces. To this end, two components of the transverse fields (e.g., $E_x$ and $E_y$) should be considered, especially when the metasurface modifies the polarization, and the 2D interpolation function $W(x,y)$ will be the impulse response of a 2D low-pass reconstruction filter. For example, for metasurfaces with rectangular lattice, this function can be selected as a 2D sinc function as $W(x,y)=\mathrm{sinc}(x/\mathrm{\Lambda_x})\mathrm{sinc}(y/\mathrm{\Lambda_y})$, which is the impulse response of an ideal 2D low-pass filter that is equal to 1 for spatial frequencies of $|k_{x,y}|\leq n_2k_0$, and 0 otherwise. Hence, the DSIR of a 2D metasurface can be similarly obtained by exciting the 2D metasurface using excitation in the form of $W(x,y)$ and sampling the fields at the lattice sites on the output reference plane. Note that the DSIR for 2D metasurfaces is a matrix and the metasurface is described by a 3D DSIR array (instead of the DSIR matrix for 1D metasurfaces).

This approach can also be extended to diffractive metasurfaces which have a lattice with a period $\mathrm{\Lambda}>\lambda/2$. For such metasurfaces, the fields are sampled at multiple points per unit cell and the sampling rate can be chosen such that the fields are sampled at an integer number of points per unit cell.

Furthermore, this approach can be used to model metasurfaces in simulation and design tools based on ray optics. Rays can be considered as beams (e.g., Gaussian beams) in the physical optics picture. Using the DSIR approach, we can find the response of a metasurface to beams and determine the outgoing wave that can be expanded in terms of outgoing beams. This leads to a ray optics model for the metasurface where a ray incident at any point of the metasurface is split into multiple rays (i.e., a non-sequential model similar to the one used to model multi-order gratings). Hence, this approach provides models for metasurfaces that can be incorporated into the ray and physical optics simulation and design tools.

\section*{Conclusion}

Analysis and design of metasurfaces require accurate methods for modeling their interactions with optical and electromagnetic waves. The DSIR concept provides a rigorous foundation for the development of simplified models of linear metasurfaces. The DSIR matrix completely and accurately characterizes a metasurface and can be used as a black-box system-level model for incorporating metasurface into ray and physical optics simulation and design tools. Although we described the concept using 1D non-diffractive metasurfaces, the DSIR concept can be readily extended to linear 2D and diffractive metasurfaces. The  two approximations we introduced significantly reduce the computational cost of DSIR enabling the design of efficient and novel metasurface based components and systems.

\section*{Methods}
The transmission coefficients of periodic structures for the conventional approach (results presented in Fig.~\ref{fig:FigThreeLabel}a) were found using the RCWA technique \citep{Liu2012}. The transmission coefficients of periodic structures composed of the same meta-atoms were found for normally incident TM plane waves. The convergence of the results was studied by increasing the number of  harmonics.

To obtain the response of aperiodic structures using the conventional approach (results shown in Figs.~\ref{fig:FigFiveLabel}b, \ref{fig:FigFiveLabel}d, and \ref{fig:FigSixLabel}d) the transmission coefficients of the meta-atoms shown in Fig.~\ref{fig:FigThreeLabel}a were used as approximations for the local transmission coefficient of the structure. The field of the incident light on the input reference plane was sampled at the lattice points and the transmitted field samples on the output reference plane (at $z=0$)  were found by multiplying the sampled field values and the local transmission coefficient at each lattice site. The fields for $z>0$ (i.e., in the region above the metasurfaces) were obtained using the plane wave expansion technique \citep{28-born2013principles} and the sampled field values on the $z=0$ plane.

The full-wave simulation results presented in Figs.~\ref{fig:FigThreeLabel}b, \ref{fig:FigFourLabel}e, \ref{fig:FigFiveLabel}b, \ref{fig:FigFiveLabel}d and \ref{fig:FigSixLabel}d were obtained using the COMSOL Multiphysics software. The metasurfaces were excited by a surface electric current density placed on the input reference plane and were surrounded by a PML layer with a thickness of 3.5 $\upmu$m.

To obtain the DSIR results of the beam deflector shown in Fig.~\ref{fig:FigFourLabel}, a metasurface with a width of 40 $\upmu$m (80 meta-atoms) was considered. Four full-wave simulations with four different surface current densities $\vec{J_\mathrm{s}}=-2\mathrm{sinc}(\frac{x-m\mathrm{\Lambda}}{\mathrm{\Lambda}})\hat{x}, m=0,1,2,3$ were used to excite the metasurface. The resulting output magnetic fields were sampled every 10 nm  on a plane 0.5 $\upmu$m above the metasurface. To remove the evanescent components, the sampled field values were filtered using an ideal low-pass filter with a cutoff frequency of $k_0$, resampled at 0.5 $\upmu$m (i.e., at lattice sites), and back-propagated to the output reference plane (on top of the meta-atoms). The responses of other meta-atoms were found by shifting these four responses to their locations. The shifted responses were used to form the DSIR matrix for the beam deflector.

The DSIR response of the beam deflector to a TM normally incident Gaussian beam (Fig.~\ref{fig:FigFourLabel}d) was found by the matrix multiplication of the DSIR matrix and the vector of sampled field values of the incident Gaussian beam on the input reference plane. The plane wave expansion technique \citep{28-born2013principles} was used to obtain the DSIR result shown in Fig.~\ref{fig:FigFourLabel}e from the sampled field values shown in Fig.~\ref{fig:FigFourLabel}d. A similar approach was used to obtain the  locality approximation results of the beam deflector (Figs.~\ref{fig:FigSixLabel}c and \ref{fig:FigSixLabel}d). The only difference is the number of meta-atoms considered in each simulation (3 to 21 meta-atoms for the results shown in Figs.~\ref{fig:FigSixLabel}c and ~\ref{fig:FigSixLabel}d versus 80 meta-atoms in Figs.~\ref{fig:FigFourLabel}d and \ref{fig:FigFourLabel}e).

The approximate DSIR matrix of the meta-atoms  in the local periodicity approximation (Fig.~\ref{fig:FigFiveLabel}b) was obtained using RCWA simulations followed by Fourier transforms. For every meta-atom with different width, a periodic structure with the same meta-atoms was considered and its transmission coefficient for TM plane-waves with incident angles ranging from -90$^\circ$ to 90$^\circ$ was found. The DSIR of the meta-atom was then computed using the Fourier transform relation between the DSIR and the angular transmission spectrum presented in the Supplementary Note 1. The DSIR matrix of the structure was generated using the DSIR responses of all the meta-atoms of different width. The responses of the metalenses to normal and oblique incident TM Gaussian beams presented in Figs.~\ref{fig:FigFiveLabel}b, 5d and 5e were found by multiplying  their DSIR matrices and the sampled input field on the input reference plane.


\begin{thebibliography}{10}
\expandafter\ifx\csname url\endcsname\relax
  \def\url#1{\texttt{#1}}\fi
\expandafter\ifx\csname urlprefix\endcsname\relax\def\urlprefix{URL }\fi
\providecommand{\bibinfo}[2]{#2}
\providecommand{\eprint}[2][]{\url{#2}}

  
\bibitem{1-Neshev2018}
\bibinfo{author}{Neshev, D.} \& \bibinfo{author}{Aharonovich, I.}
\newblock \bibinfo{title}{Optical metasurfaces: new generation building blocks for multi-functional optics}.
\newblock \emph{\bibinfo{journal}{Light: Sci. Appl.}}
 \textbf{\bibinfo{volume}{7}},
  \bibinfo{pages}{1-5} (\bibinfo{year}{2018}).
  
\bibitem{2-Kamali2018}
\bibinfo{author}{Kamali, S. M.}, \bibinfo{author}{Arbabi, E.}
\bibinfo{author}{Arbabi, A.} \&
 \bibinfo{author}{Faraon, A.}
\newblock \bibinfo{title}{A review of dielectric optical metasurfaces for wavefront control}.
\newblock \emph{\bibinfo{journal}{Nanophotonics}} \textbf{\bibinfo{volume}{7}},
  \bibinfo{pages}{1041-1068} (\bibinfo{year}{2018}).
    
\bibitem{3-Chen2016}
\bibinfo{author}{Chen, H.-T.}, \bibinfo{author}{Taylor, A. J.} \&
 \bibinfo{author}{Yu, N.}
\newblock \bibinfo{title}{A review of metasurfaces: physics and applications}.
\newblock \emph{\bibinfo{journal}{Rep. Progress. Phys}} \textbf{\bibinfo{volume}{79}},
  \bibinfo{pages}{076401} (\bibinfo{year}{2016}).    
 
\bibitem{4-Chang2018}
\bibinfo{author}{Chang, S.}, \bibinfo{author}{Guo, X.} \&
\bibinfo{author}{Ni, X.}
\newblock \bibinfo{title}{Optical metasurfaces: progress and applications}.
\newblock \emph{\bibinfo{journal}{Annu. Rev. Mater. Res.}}			
  \bibinfo{pages}{279-302} (\bibinfo{year}{2018}). 
  
\bibitem{5-Scheuer2017}
\bibinfo{author}{Scheuer, J.}
\newblock \bibinfo{title}{Metasurfaces-based holography and beam shaping: engineering the phase profile of light}.
\newblock \emph{\bibinfo{journal}{Nanophotonics}} \textbf{\bibinfo{volume}{6}},
  \bibinfo{pages}{137–152} (\bibinfo{year}{2016}).    
  
\bibitem{6-Jahani2016}
\bibinfo{author}{Jahani, S.} \& \bibinfo{author}{Jacob, Z.}
\newblock \bibinfo{title}{All-dielectric metamaterials}.
\newblock \emph{\bibinfo{journal}{Nat. Nanotechnol.}} \textbf{\bibinfo{volume}{11}},
  \bibinfo{pages}{23–36} (\bibinfo{year}{2016}).  
  
\bibitem{7-Kildishev2013}
\bibinfo{author}{Kildishev, A. V.}, \bibinfo{author}{Boltasseva, A.} \&
 \bibinfo{author}{Shalaev, V. M.}
\newblock \bibinfo{title}{Planar photonics with metasurfaces}.
\newblock \emph{\bibinfo{journal}{Science}} \textbf{\bibinfo{volume}{339}},
  \bibinfo{pages}{1232009} (\bibinfo{year}{2013}).    
  
\bibitem{8-Zhan2017}
\bibinfo{author}{Zhan, A.}, \bibinfo{author}{Colburn, S.}
\bibinfo{author}{Dodson, C. M.} \&
 \bibinfo{author}{Majumdar, A.}
\newblock \bibinfo{title}{Metasurface freeform nanophotonics}.
\newblock \emph{\bibinfo{journal}{Sci. Rep.}} \textbf{\bibinfo{volume}{7}},
  \bibinfo{pages}{1673} (\bibinfo{year}{2017}).  
  
\bibitem{9-Liu2018}
\bibinfo{author}{Liu, W.}, \bibinfo{author}{Zhang, Y.}
\bibinfo{author}{Gao, J.}, \&
 \bibinfo{author}{Yang, X.}
\newblock \bibinfo{title}{Generation of three-dimensional optical cusp beams with ultrathin metasurfaces}.
\newblock \emph{\bibinfo{journal}{Sci. Rep.}} \textbf{\bibinfo{volume}{8}},
  \bibinfo{pages}{9493} (\bibinfo{year}{2018}). 

\bibitem{10-Horie2018}
\bibinfo{author}{Jang, M.} \emph{et~al.}
\newblock \bibinfo{title}{Wavefront shaping with disorder-engineered metasurfaces}.
\newblock \emph{\bibinfo{journal}{Nat. Photon.}} \textbf{\bibinfo{volume}{12}},
  \bibinfo{pages}{84-90} (\bibinfo{year}{2018}). 

\bibitem{11-Khorasaninejad2017}
\bibinfo{author}{Khorasaninejad, M.}, \& \bibinfo{author}{Capasso, F.}
\newblock \bibinfo{title}{Metalenses: Versatile multifunctional photonic components}.
\newblock \emph{\bibinfo{journal}{Science}} \textbf{\bibinfo{volume}{358}},
  \bibinfo{pages}{eaam8100} (\bibinfo{year}{2017}). 

\bibitem{12-Ang2017}
\bibinfo{author}{ Wang, L.} \emph{et~al.}
\newblock \bibinfo{title}{Grayscale transparent metasurface holograms}.
\newblock \emph{\bibinfo{journal}{Optica}} \textbf{\bibinfo{volume}{3}},
  \bibinfo{pages}{1504-1505} (\bibinfo{year}{2016}). 
  
\bibitem{13-Aieta2012}
\bibinfo{author}{Aieta, F.} \emph{et~al.}
\newblock \bibinfo{title}{Aberration-free ultrathin flat lenses and axicons at telecom wavelengths based on plasmonic metasurfaces}.
\newblock \emph{\bibinfo{journal}{Nano Lett.}} \textbf{\bibinfo{volume}{12}},
  \bibinfo{pages}{4932-4936} (\bibinfo{year}{2012}).  
  
\bibitem{14-Arbabi2015}
\bibinfo{author}{Arbabi, A.}, \bibinfo{author}{Horie, Y.}
\bibinfo{author}{Ball, A. J.}, \bibinfo{author}{Bagheri, M.} \&
 \bibinfo{author}{Faraon, A.}
\newblock \bibinfo{title}{Subwavelength-thick
lenses with high numerical apertures and large efficiency based on highcontrast transmitarrays}.
\newblock \emph{\bibinfo{journal}{Nat. Commun.}} \textbf{\bibinfo{volume}{6}},
  \bibinfo{pages}{7069} (\bibinfo{year}{2015}).  
    
\bibitem{15-Arbabi2015b}
\bibinfo{author}{Arbabi, A.}, \bibinfo{author}{Horie, Y.}
\bibinfo{author}{Bagheri, M.}, \&
 \bibinfo{author}{Faraon, A.}
\newblock \bibinfo{title}{Dielectric metasurfaces for
complete control of phase and polarization with subwavelength spatial
resolution and high transmission}.
\newblock \emph{\bibinfo{journal}{Nat. Nanotechnol.}} \textbf{\bibinfo{volume}{10}},
\bibinfo{pages}{937–943} (\bibinfo{year}{2015}).  

\bibitem{16-Vo2014}
\bibinfo{author}{Vo, S.} \emph{et~al.}
\newblock \bibinfo{title}{Sub-wavelength grating lenses with a twist}.
\newblock \emph{\bibinfo{journal}{IEEE Photon. Technol. Lett.}} \textbf{\bibinfo{volume}{26}},
\bibinfo{pages}{1375–1378} (\bibinfo{year}{2014}).  

\bibitem{17-Devlin2016}
\bibinfo{author}{Khorasaninejad, M.} \emph{et~al.}
\newblock \bibinfo{title}{Metalenses at visible wavelengths: Diffraction-limited focusing and subwavelength resolution imaging}.
\newblock \emph{\bibinfo{journal}{Science}} \textbf{\bibinfo{volume}{352}},
  \bibinfo{pages}{1190–1194} (\bibinfo{year}{2016}).  
  
\bibitem{18-Arbabi2016}
\bibinfo{author}{Arbabi, A.} \emph{et~al.}
\newblock \bibinfo{title}{Miniature optical planar camera based on a wide-angle metasurface doublet corrected for monochromatic aberrations}.
\newblock \emph{\bibinfo{journal}{Nat. Commun.}} \textbf{\bibinfo{volume}{7}},
  \bibinfo{pages}{13682} (\bibinfo{year}{2016}).
  
\bibitem{19-Arbabi2017a}
\bibinfo{author}{Arbabi, A.}, \bibinfo{author}{Arbabi, E.}
\bibinfo{author}{Horie, Y.}, \bibinfo{author}{Kamali, S. M.} \&
 \bibinfo{author}{Faraon, A.}
\newblock \bibinfo{title}{Planar metasurface retroreflector}.
\newblock \emph{\bibinfo{journal}{Nat. Photon.}} \textbf{\bibinfo{volume}{11}},
  \bibinfo{pages}{415–420} (\bibinfo{year}{2017}).
  
\bibitem{20-Arbabi2017b}
\bibinfo{author}{Arbabi, E.}, \bibinfo{author}{Arbabi, A.}
\bibinfo{author}{Kamali, S. M.}, \bibinfo{author}{Horie, Y.} \&
 \bibinfo{author}{Faraon, A.}
\newblock \bibinfo{title}{Controlling the sign of chromatic dispersion in diffractive optics with dielectric metasurfaces}.
\newblock \emph{\bibinfo{journal}{Optica}} \textbf{\bibinfo{volume}{4}},
  \bibinfo{pages}{625-632} (\bibinfo{year}{2017}).
  
\bibitem{21-Zhao2015}
\bibinfo{author}{Zhao, Z.} \emph{et~al.}
\newblock \bibinfo{title}{Multispectral optical metasurfaces enabled by achromatic phase
transition}.
\newblock \emph{\bibinfo{journal}{Sci. Rep.}} \textbf{\bibinfo{volume}{5}},
  \bibinfo{pages}{15781} (\bibinfo{year}{2015}).
  
\bibitem{22-Yuan2017}
\bibinfo{author}{Yuan, J.}, \bibinfo{author}{Yin, G.}
\bibinfo{author}{Jiang, W.}, \bibinfo{author}{Wu, W.} \&
 \bibinfo{author}{Ma, Y.}
\newblock \bibinfo{title}{Design of mechanically robust metasurface lenses for RGB colors}.
\newblock \emph{\bibinfo{journal}{J. Optics}} \textbf{\bibinfo{volume}{19}},
  \bibinfo{pages}{105002} (\bibinfo{year}{2017}).

\bibitem{23-Kruk2016}
\bibinfo{author}{Kruk, S.} \emph{et~al.}
\newblock \bibinfo{title}{Broadband highly efficient dielectric metadevices for polarization control}.
\newblock \emph{\bibinfo{journal}{APL Photonics}} \textbf{\bibinfo{volume}{1}},
  \bibinfo{pages}{030801} (\bibinfo{year}{2016}).

\bibitem{24-Pfeiffer2013}
\bibinfo{author}{Pfeiffer, C.} \& \bibinfo{author}{Grbic, A}
\newblock \bibinfo{title}{Cascaded metasurfaces for complete phase and polarization control}.
\newblock \emph{\bibinfo{journal}{Appl. Phys. Lett.}} \textbf{\bibinfo{volume}{102}},
  \bibinfo{pages}{231116} (\bibinfo{year}{2013}).

\bibitem{25-Yang2014}
\bibinfo{author}{Yang, Y.} \emph{et~al.}
\newblock \bibinfo{title}{Dielectric meta-reflectarray for broadband linear polarization conversion and optical vortex generation}.
\newblock \emph{\bibinfo{journal}{Nano lett.}} \textbf{\bibinfo{volume}{14}},
  \bibinfo{pages}{1394-1399} (\bibinfo{year}{2014}).

\bibitem{26-harrington1961time}
\bibinfo{author}{Harrington, R. F.} 
\newblock \emph{\bibinfo{title}{Time-Harmonic Electromagnetic Fields}}
  (\bibinfo{publisher}{McGraw-Hill},
   \bibinfo{year}{1961}).

\bibitem{27-mersereau1984multidimensional}
\bibinfo{author}{Dudgeon, D. E.} \& \bibinfo{author}{Russell M. M.}
\newblock \emph{\bibinfo{title}{Multidimensional digital signal processing}}
 (\bibinfo{publisher}{Prentice-Hall},
   \bibinfo{year}{1984}).

\bibitem{28-born2013principles}
\bibinfo{author}{Born, M.} \& \bibinfo{author}{Wolf, E.}
\newblock \emph{\bibinfo{title}{Principles of optics}}
  (\bibinfo{publisher}{Cambridge University Press},
  \bibinfo{year}{1999}).

\bibitem{29-moharam1981rigorous}
\bibinfo{author}{Moharam, M. G.} \& \bibinfo{author}{Gaylord, T. K.}
\newblock \bibinfo{title}{Rigorous coupled-wave analysis of planar-grating diffraction}.
\newblock \emph{\bibinfo{journal}{J. Opt. Soc. Am.}} \textbf{\bibinfo{volume}{71}},
  \bibinfo{pages}{811–818} (\bibinfo{year}{1981}).
  

\bibitem{Liu2012}
\bibinfo{author}{Liu, Victor} \& \bibinfo{author}{Fan, Shanhui}
\newblock \bibinfo{title}{S 4: A free electromagnetic solver for layered periodic structures}.
\newblock \emph{\bibinfo{journal}{Computer Physics Communications}}
\textbf{\bibinfo{volume}{183}},
  \bibinfo{pages}{2233-2244} (\bibinfo{year}{2012}).


\end{thebibliography}

\begin{thebibliography}{10}
\expandafter\ifx\csname url\endcsname\relax
  \def\url#1{\texttt{#1}}\fi
\expandafter\ifx\csname urlprefix\endcsname\relax\def\urlprefix{URL }\fi
\providecommand{\bibinfo}[2]{#2}
\providecommand{\eprint}[2][]{\url{#2}}
\bibitem{30-oppenheim1999discrete}
\bibinfo{author}{Oppenheim, A. V.} \& \bibinfo{author}{Schafer, R. W}
\newblock \emph{\bibinfo{title}{Discrete-Time Signal Processing}}
  (\bibinfo{publisher}{Pearson Education India},
   \bibinfo{year}{1999}).
   
\end{thebibliography}

\bibliographystyle{plain}

\vspace{0.2in}
\section*{Acknowledgements} 
This work was supported by the DARPA Extreme Optics and Imaging program.

\setcounter{equation}{0}
\newpage
\maketitle
\section*{Supplementary Note 1: Relation between DSIR and angular transmission spectrum of non-diffractive periodic metasurfaces}
Here we show that there is a Fourier transform relation between the DSIR of a non-diffractive periodic metasurface and its transmission angular spectrum. The Fourier transform of a discrete signal $t(n)$ is defined as \cite{30-oppenheim1999discrete}
\begin{equation}
\widetilde{t}(\omega)=\sum_{n=-\infty}^{\infty}t(n)\mathrm{e}^{j\omega n},
\end{equation}
and the inverse transform is given by
\begin{equation}
t(n)=\frac{1}{2\pi}\int_{-\pi}^{\pi}\widetilde{t}(\omega)\mathrm{e}^{-j\omega n}\, \mathrm{d}\omega.
\end{equation}
To find the DSIR, the metasurface is excited by an incident wave whose field on the input reference plane is given by $F_{\mathrm{in}}(x)=\mathrm{sinc}(x/\Lambda)$. The incident wave can be expressed as a sum of plane waves. To find the outgoing wave we find the response of the metasurface to these plane waves and superimpose them to form the outgoing wave. Using the Fourier transform of a sinc function, we can write
\begin{equation}
F_{\mathrm{in}}(x)=\mathrm{sinc}(\frac{x}{\mathrm{\Lambda}})=\frac{\mathrm{\Lambda}}{2\pi}\int_{-\pi/\Lambda}^{\pi/\Lambda}\mathrm{e}^{-jk_xx}\, \mathrm{d}k_x,
\end{equation}
which is a continuous sum over plane waves $\mathrm{e}^{-jk_xx}$ that are incident at angle $\theta$ where $\mathrm{sin}(\theta)=\frac{k_x}{n_1k_0}$.  Let $T(\theta)$ represents the complex-valued transmission coefficient of the metasurface for a plane wave incident at angle $\theta$ (i.e., the ratio of the tangential component of the transmitted field on output reference plane to that of the incident one on the input reference plane). The outgoing field can be written as a sum of plane waves whose amplitudes are modified by the transmission coefficient
\begin{equation}
F_{\mathrm{out}}(x)=\frac{\mathrm{\Lambda}}{2\pi}\int_{-\pi/\Lambda}^{\pi/\Lambda}T(\theta)\mathrm{e}^{-jk_xx}\, \mathrm{d}k_x.
\end{equation}
The DSIR is obtained by sampling the outgoing wave 
\begin{equation}
t(n)=F_{\mathrm{out}}(n\mathrm{\Lambda})=\frac{\mathrm{\Lambda}}{2\pi}\int_{-\pi/\Lambda}^{\pi/\Lambda}T(\theta)\mathrm{e}^{-jk_xn\Lambda}\, \mathrm{d}k_x=\frac{1}{2\pi}\int_{-\pi}^{\pi}T(\theta)\mathrm{e}^{-j\omega n}\, \mathrm{d}\omega
\end{equation}
where we have defined $\omega=k_x\Lambda$. Comparing (5) and the inverse Fourier transform relation (2), we obtain\\
\begin{equation}
T(\theta)=\widetilde{t}(\omega)=\widetilde{t}(k_x\Lambda)=\widetilde{t}(n_1k_0  \mathrm{ sin}(\theta) \mathrm{\Lambda})=\widetilde{t}(2\pi\mathrm{sin}(\theta)\frac{\mathrm{\Lambda}}{\lambda_1})
\end{equation}

\section*{Supplementary Note 2: Bound on the Euclidean norm of DSIR of non-diffractive periodic metasurfaces}
Here we show that the Euclidean ($\mathrm{L^2}$) norm of the DSIR of a non-diffractive periodic metasurface is bounded by one. According to the Parseval's theorem
\begin{equation}
\sum_{n=-\infty}^{\infty}|t(n)|^2=\frac{1}{2\pi}\int_{\mathrm{-\pi}}^{\mathrm{\pi}}|\widetilde{t}(\omega)|^2\mathrm{d}\omega
\end{equation}
According to (6) $\widetilde{t}(\omega)=T(\theta)$, and for passive metasurfaces the transmission coefficient modulus is smaller than unity, thus 
\begin{equation}
\sum_{n=-\infty}^{\infty}|t(n)|^2=\frac{1}{2\pi}\int_{\mathrm{-\pi}}^{\mathrm{\pi}}|\widetilde{t}(\omega)|^2\mathrm{d}\omega<=\frac{1}{2\pi}\int_{\mathrm{-\pi}}^{\mathrm{\pi}}\mathrm{d}\omega=1
\end{equation}

\bibliographystyle{plain}

\end{document}